\documentclass{aarw}
\usepackage{graphicx}
\usepackage{txfonts}
\usepackage{natbib}
\bibpunct{(}{)}{;}{a}{}{,} % to follow A&A style
\begin{document}
\title{A model for delayed emission in a very-high energy gamma-ray flare in \object{Markarian 501}}
\titlerunning{Delayed emission in a VHE $\gamma$-ray flare in \object{Mrk 501}}
\author{W.~Bednarek\inst{1} \and R.~M.~Wagner\inst{2}}
\authorrunning{Bednarek \and Wagner}

\institute{Department of Experimental Physics,
 University of \L\'od\'z, ul. Pomorska 149/153, 90-236 \L\'od\'z, Poland\\
 \email{bednar@fizwe4.fic.uni.lodz.pl}
\and
 Max-Planck-Institut f\"ur Physik, F\"ohringer Ring 6, D-80805 M\"unchen, Germany\\
 \email{robert.wagner@mppmu.mpg.de}
}

%\date{Received 15 November 2007; Accepted 9 May 2008; Preprint online version 3 June 2008}
\date{Received 15 November 2007; Accepted 9 May 2008}

\abstract
% context
{Recently, the MAGIC collaboration reported evidence for a delay in
the arrival times of photons of different energies during a
$\gamma$-ray flare from the blazar Markarian 501 on 2005 July 9.}
% aims
{We aim at describing the observed delayed high-energy emission.}
% methods
{We apply a homogeneous synchrotron self-Compton (SSC) model under
the assumption that the blob containing relativistic electrons was
observed in its acceleration phase.}
% results
{This modified SSC model predicts the appearance of a
$\gamma$-ray flare first at lower energies and subsequently at
higher energies. }
% conclusions
{Based on the reported time delay of $\sim 240 \,\mbox{s}$ between
the flare observed at $\sim 190 \,\mbox{GeV}$ and $2.7
\,\mbox{TeV}$, we predict a delay on the order of
$1\,\mbox{h}$ if observed between $10 \,\mbox{GeV}$ and
$100 \,\mbox{GeV}$. Such delay timescales can be tested in the
future by simultaneous flare observations with the Gamma Ray 
Large Area Space Telescope (GLAST) and Cherenkov telescopes.}


\maketitle

\section{Introduction}
Recent observations of energy-dependent, few-minute
timescale TeV $\gamma$-ray flares in the blazar \object{Markarian
(Mrk) 501}~\citep{alb07a} are difficult to explain in terms of the
``classical'' homogeneous synchrotron self-Compton (SSC) model
\citep[e.g.,][]{mg85,mgc92} in which the emission region is
moving along the jet with a constant Lorentz factor $\sim 10-20$
\citep{g04}, as inferred from the observations of superluminal
motion in active galactic nuclei. Similar few-minute
timescale flares have also been observed recently from
\object{PKS 2155-304}~\citep{aha07}. However, in the case of
\object{PKS 2155-304}, no energy-dependent flare evolution was
found.

Homogeneous blob models will require blobs moving with
Lorentz factors significantly larger than mentioned above, see,
e.g., Fig.~3c in \Citet{bp99}, or the more recent papers by
\citet{bfr07} and ~\citet{aha07}. In the case of the 2005 July 9 flare of
\object{Mrk 501}, an arrival time delay of $239\pm 78 \,\mbox{s}$
was observed between $\gamma$-rays in the  $>1.2 \,\mbox{TeV}$
energy range with respect to the $0.15 - 0.25 \,\mbox{TeV}$ energy
range \citep{alb07a}. This delay was confirmed by a more detailed
quantitative investigation with an unbinned, photon-by-photon
analysis~\citep{alb07b}. 

Two solutions for these fine emission features have been
considered up to now, viz. gradual acceleration of electrons
inside the blob~\citep{alb07a} and the attribution of the
delay to a source-extrinsic $\gamma$-ray transport
effect~\citep{alb07b}. In the gradual-acceleration scenario,
particles inside the emission region moving with constant Doppler
factor need some time to be accelerated to energies
that allow them to produce $\gamma$-rays with specific
energies. Therefore, a $\gamma$-ray flare should first appear at lower
energies and only later on at larger energies. The
second scenario relates the delay between different
energies to the energy-dependent speed of photons in vacuum.
The effect that lower-energy photons arrive before the
higher-energy ones is predicted
in some models of quantum-gravity 
\citep[see][for details]{alb07b}.

The large Doppler factors inferred from TeV $\gamma$-ray observations
have to be achieved somewhere in the very inner
parts of the jet. Therefore, it is natural to expect that the
emission region has to undergo an initial acceleration phase. 
Very-Large Baseline Interferometry (VLBI) observations rather show 
evidence for decelerating radio-emitting blobs \citep[e.g.,][for Mrk~501]{ep02}.
However, the very inner regions where we assume the initial accereration to
take place, 
could not be investigated yet with VLBI.

We show that a delayed emission of the higher-energy
$\gamma$-rays can be naturally explained by assuming that the
blob accelerates significantly during its propagation along the
jet, producing more energetic $\gamma$-rays when the Lorentz
factor of the blob is larger. We argue that for the first time
$\gamma$-ray emission in BL~Lac objects has been captured in the
initial acceleration phase of the blob.

\section{Accelerating blob model}
We consider a blob of relativistic electrons that moves along the
jet with a Lorentz factor $\gamma_{\rm b}$ at any specific moment.
We assume a ``classical'' SSC model, in which electrons are
accelerated inside the blob by shock acceleration, with an
acceleration efficiency proportional to the magnetic field
strength at the location of the blob. Due to the presence of
magnetic fields, the electrons inside the blob lose energy by
synchrotron emission. This soft radiation is upscattered to  
$\gamma$-ray energies by the same population of electrons. In
terms of this classical picture, X-ray and TeV $\gamma$-ray flares
are naturally explained. However, the observation of a time delay
between $\gamma$-rays of different energies belonging to the same
flare cannot be easily understood in such a simple scenario.
Therefore, we modify the homogeneous SSC model by assuming that
the blob has been captured during its acceleration phase in the
inner part of the jet. Let us consider the blob acceleration
between its Lorentz factors $\gamma_{\rm b}^{\rm min}$ and
$\gamma_{\rm b}^{\rm max}$. We assume that the magnetic field
strength in the jet (and also in the blob) drops proportionally to
the distance from the base of the jet, $B\propto X$,
which is a good approximation in the case of a conical jet with a
fixed opening angle at some distance to the base of the jet.
In our model, we discuss a quasi-spherical blob with a radius
\begin{eqnarray}
R_{\rm b} \propto X,
\label{eq0}
\end{eqnarray}
\noindent
i.e., proportional to its distance from the
base of the jet. This is expected for a conical jet, whose borders
limit the expansion of the blob.

We propose that a delay observed in a flare between different energy
ranges is due to the fact that the lower-energy part of the flare
is produced when the blob is closer to the base of the jet (moving
with a lower Lorentz factor), while the higher-energy part is
produced at larger distances from the jet base, at which the blob
has accelerated to a higher Lorentz factor.

We now derive a description of the acceleration of the blob along
the jet. Observations show that the TeV $\gamma$-ray luminosities
found for \object{Mrk 501} during the 2005 July 9 flare in the
above-mentioned two energy ranges are comparable~\citep{alb07a}.
The observed $\gamma$-ray luminosity is related to the luminosity
in the blob frame by
\begin{eqnarray}
\dot{P}_{\rm obs}^{\rm IC}\propto \dot{P}_{\rm b}^{\rm IC} D_{\rm
b}^2, \label{eq1}
\end{eqnarray}
\noindent where $D_{\rm b}$ is the Doppler factor of the blob. In
the simplest possible case, in which the observer is located on
the axis of the jet, the blob Doppler factor is related to its
Lorentz factor by $D_{\rm b} = [\gamma_{\rm b}(1 - \beta_{\rm
b}\cos\theta)]^{-1}\simeq 2\gamma_{\rm b}$, where $\cos\theta = 1$
is the cosine of the observation angle, assumed $\theta = 0\degr$,
and $\beta_{\rm b} = v_{\rm b}/c$ is the velocity of the blob.
Moreover, the $\gamma$-ray luminosity in the blob frame depends on
the energy density of the synchrotron radiation and the Lorentz
factor of the electrons,
\begin{eqnarray}
\dot{P}_{\rm b}^{\rm IC}\propto U_{\rm syn}\gamma_{\rm e}^2N_{\rm
e}, \label{eq2}
\end{eqnarray}
\noindent where $\gamma_{\rm e}$ is the Lorentz factor of the
electrons that produce $\gamma$-rays (with characteristic
energies for which a time delay has been reported), and $N_{\rm
e}$ is the number of relativistic electrons inside the blob, which
is assumed to be independent of time. We calculate the
average energy density of synchrotron photons inside the blob by
estimating the synchrotron energy losses of the complete electron
population inside the blob. That is, the energy losses of all
electrons inside the blob multiplied by the average time that the
synchrotron photons spend inside the blob, $\sim R_b/c$,
divided by the volume of the blob,
\begin{eqnarray}
U_{\rm syn}\propto \frac{\dot{P}_{\rm syn}}{V_{\rm b}}\frac{R_{\rm
b}}{c}, \label{eq2a}
\end{eqnarray}
\noindent
where
\begin{eqnarray}
\dot{P}_{\rm syn}\propto B^2\gamma_{\rm e}^2N_{\rm e}
\label{eq2b}
\end{eqnarray}
\noindent are the synchrotron energy losses of electrons, $R_{\rm b}$ is the
radius of the blob, and $V_{\rm b}\propto R_{\rm b}^3$ is the volume of the
quasi-spherical blob. Provided that the energies of the electrons inside the
blob are limited by their synchrotron energy losses, we compare their
synchrotron energy losses with their energy gains from the acceleration
mechanism, which scales as $\dot{P}_{\rm acc}\propto B$ in the shock
acceleration scenario.  Then the maximum energy of the electrons is 
\begin{eqnarray}
\gamma_{\rm e}\propto B^{-1/2}.
\label{eq3}
\end{eqnarray}
\noindent In this case the synchrotron energy losses of the most energetic
electrons are only linearly proportional to the magnetic field
(Eq.~\ref{eq2b}). The blob is assumed to expand during its propagation along
the jet proportionally to the distance traversed, i.e. its volume
increases like $V_{\rm b}\propto X^3$. From these simple
considerations we conclude that the $\gamma$-ray luminosity in the blob frame
depends on the distance from the base of the jet as
\begin{eqnarray}
\dot{P}_{\rm b}^{\rm IC}\propto R_{\rm b}^{-2}. \label{eq4}
\end{eqnarray}
\noindent Since the observed $\gamma$-ray luminosities
($\dot{P}_{\rm obs}^{\rm IC}$) in the considered energy range are
approximately constant, the Doppler factor of the blob has to
increase during the propagation of the blob along the jet like
$D_{\rm b}\propto R_{\rm b}$ to keep the $\gamma$-ray
luminosity approximately constant during the flare (cf.
Eq.~\ref{eq1}). Since the blob moves with a large Lorentz factor,
the traversed distance is given approximately by $X\simeq c t
\propto R_{\rm b}$, i.e., its Lorentz factor has to be
approximately proportional to the time during which it travels
inside the jet. Note that the time $t$ is measured in the
stationary jet frame. Thus,
\begin{eqnarray}
\gamma_{\rm b}\simeq 0.5D_{\rm b} = A t,
\label{eq5}
\end{eqnarray}
\noindent where $A$ is a constant. We conclude that the observed time delay
between different energies can be related to the acceleration phase of the blob
inside the jet. In the next section, this conclusion is further investigated to
constrain the blob acceleration scenario for the case of the recent
observations of \object{Mrk 501}~\citep{alb07a}.

\section{Energy-delayed $\gamma$-ray flare}
In our model, electrons are accelerated to maximum energies that
change during the propagation of the blob due to a change of the
local magnetic field strength in the jet. Thus, the $\gamma$-rays
produced by these electrons also might have characteristic
energies that are related to the energies of their parent electrons.
The energy of a produced $\gamma$-ray photon can be estimated by
\begin{eqnarray}
E_{\gamma}\simeq m_{\rm e}\gamma_{\rm e}D_{\rm b}.
\label{eq5b}
\end{eqnarray}
\noindent This estimate is reasonable in the case of a relatively
flat power-law spectrum of synchrotron photons (as observed during
the \object{Mrk 501} flares), i.e., when the inverse Compton
scattering mainly occurs in the transition region between the Thomson and the
Klein-Nishina regime. The time delay between the appearance of the
flare at two different energies, $E_\gamma^{\rm min}$ and
$E_\gamma^{\rm max}$, is due to the production of the
corresponding $\gamma$-rays at different stages of the
acceleration of the blob. While a photon with energy
$E_\gamma^{\rm min}$ moves the distance $c\,{\mathrm d}t$, the
blob traverses a distance $\beta c {\mathrm d}t$, where ${\mathrm d}t$
is the time measured in the jet frame, not in the blob frame.
Therefore, the distance (in the jet frame) between two photons
$E_\gamma^{\rm min}$ and $E_\gamma^{\rm max}$ is ${\mathrm
d}(\Delta X) = c(1 - \beta){\mathrm d}t$. The distance between
these photons corresponds to their time delay measured in the
observer frame ${\mathrm d}(\Delta\tau) = {\mathrm d}(\Delta
X)/c$. The total time delay in the case of a blob with changing
velocity can be calculated as
\begin{eqnarray}
\Delta\tau = \int_{t_{\rm min}}^{t_{\rm max}} \left(1 -
\beta(t)\right) {\rm d}t\cong \int_{\gamma_{\rm b}^{\rm
min}}^{\gamma_{\rm b}^{\rm max}}{{{\rm d}t}\over{2\gamma_{\rm
b}^2}}, \label{eq6}
\end{eqnarray}
\noindent where $1 - \beta(t)\cong 1/2\gamma_{\rm b}^2$, $t$ is
the propagation time of the blob inside the jet, and $t_{\rm min}$ and
$t_{\rm max}$ denote the moments at which the Lorentz factor of
the blob is equal to $\gamma_{\rm b}^{\rm min}$ and $\gamma_{\rm
b}^{\rm max}$, respectively. By applying the relation between the
Lorentz factor of the blob and the travel time (Eq.~\ref{eq5}),
we arrive at a formula for the observational time delay
between $\gamma$-rays of different energies
\begin{eqnarray}
\Delta\tau\cong {{1}\over{2A}}\int_{\gamma_{\rm b}^{\rm
min}}^{\gamma_{\rm b}^{\rm max}}{{{\rm d}\gamma_{\rm
b}}\over{\gamma_{\rm b}^2}} =
{{1}\over{2A}}\left({{1}\over{\gamma_{\rm b}^{\rm
min}}}-{{1}\over{\gamma_{\rm b}^{\rm max}}}\right)\approx
{{1}\over{2A\gamma_{\rm b}^{\rm min}}}, \label{eq7}
\end{eqnarray}
\noindent provided that $\gamma_{\rm b}^{\rm max}\gg  \gamma_{\rm
b}^{\rm min}$. This simple relation for time
delays in the flare observed at different photon energies holds true 
as long as these energies are directly linked to the energies
of the parent electrons accelerated in the blob. It is obvious
that the lower the energy ranges are chosen, the larger the
observed time delay will be: For example, a time delay between
$\gamma$-rays of $30 \,\mbox{GeV}$ and $300 \,\mbox{GeV}$  should
be larger by an order of magnitude than a delay observed between
$\gamma$-rays of $300 \,\mbox{GeV}$ and $3 \,\mbox{TeV}$. This
clear prediction can be tested in the near future by simultaneous
observations in the GeV energy range (with GLAST) and the TeV
energy range (with MAGIC, H.E.S.S., and VERITAS).

During its acceleration from $\gamma_{\rm
b}^{\rm min}$ to $\gamma_{\rm b}^{\rm max}$, the blob covers a
distance
\begin{eqnarray}
X_{\rm acc} & = & c\int_{t_{\rm min}}^{t_{\rm max}} \beta(t){\rm d}t\cong {{c}\over{A}}\int_{\gamma_{\rm b}^{\rm min}}^{\gamma_{\rm b}^{\rm max}}\left(1 - 0.5\gamma_{\rm b}^{-2}\right){\rm d}\gamma_{\rm b} \nonumber \\
            & = & {{c}\over{A}} \left[\left(\gamma_{\rm b}^{\rm max} - \gamma_{\rm b}^{\rm min}\right) - {{1}\over{2}}\left(
{{1}\over{\gamma_{\rm b}^{\rm min}}} - {{1}\over{\gamma_{\rm
b}^{\rm max}}}\right)\right]  \nonumber \\
  & \approx &{{c}\over{A}}
\left(\gamma_{\rm b}^{\rm max} - \gamma_{\rm b}^{\rm min}\right)
\label{eq8}
\end{eqnarray}
\noindent in the jet.
By using Eq.~\ref{eq7}, we can express the distance $X_{\rm acc}$ traveled
by the blob by the measured time delay $\Delta\tau$ and the
limiting values of the Lorentz factors of the blob,
\begin{eqnarray}
X_{\rm acc}  =  c \Delta\tau \left(2\gamma_{\rm b}^{\rm max}\gamma_{\rm
b}^{\rm min} - 1\right). \label{eq9}
\end{eqnarray}
\section{Constraints on the Doppler factor of the blob}
Since a homogeneous SSC model is considered, we can constrain the
Doppler factor of the blob in two ways based (a) on the known
variability timescales of the synchrotron and $\gamma$-ray emission;
and (b) on the escape condition of $\gamma$-rays from the
synchrotron radiation inside the blob. The maximum energy of the
observed synchrotron photons can be estimated by
\begin{eqnarray}
\varepsilon_{\rm s}\cong m_{\rm e} \frac{B}{B_{\rm
cr}}\gamma_{\rm e}^2 D_{\rm b} \cong m_{\rm e}
\frac{B}{B_{\rm
cr}} \left(\frac{E_\gamma^{\rm max}}{m_{\rm
e}}\right)^2 D_{\rm b}^{-1}, \label{eq10}
\end{eqnarray}
\noindent where $B_{\rm cr} = 4.4\times 10^{13} \,\mbox{G}$ and
$\gamma_{\rm e}$ can be approximated by applying Eq.~\ref{eq5b},
$\gamma_{\rm e}\approx E_\gamma^{\rm max}/(m_{\rm e}D_{\rm b})$.
On the other hand, the observed emission variability timescale of
the flare has to be at least equal to (or larger than) the
synchrotron cooling time of the electrons to observe a
synchrotron flare, i.e.,
\begin{eqnarray}
\tau_{\rm var} \ge \frac{\tau_{\rm s}}{D_{\rm b}}, \label{eq11}
\end{eqnarray}
\noindent where $\tau_{\rm s} = 3 m_{\rm e}\gamma_{\rm e}/C
B^2\gamma_{\rm e}^2$, and $C$ is a constant describing the
efficiency of synchrotron energy losses. By estimating the
magnetic field strength from Eq.~\ref{eq10} and reversing
Eq.~\ref{eq11}, we obtain a lower limit on the Doppler factor of
the blob,
\begin{eqnarray}
D_{\rm b}^2\ge \frac{E_{\gamma, \rm max}^3}{C B_{\rm
cr}^2\tau_{\rm var}\varepsilon_{\rm s}^2}. \label{eq12}
\end{eqnarray}
\noindent By considering the most extreme parameters ever observed
for \object{Mrk 501}, i.e., $E_{\gamma, \rm max}\approx 20$
TeV~\citep{kon99,aha01}, $\varepsilon_{\rm s}\approx 0.5$
MeV~\citep{cat97}, and $\tau_{\rm var}\approx 200$ s
\citep{alb07a}, we estimate the lower limit on the Doppler factor
of the blob to be $D_{\rm b}\approx 10$, and the lower limit on
the magnetic field strength inside the blob region to be $B\approx
0.3 \,\mbox{G}$. Note that these values are consistent with 
estimates derived from more complete information
on a flare in \object{Mrk 501} \citep[e.g.,][]{bp99} and with
Doppler factors $\sim 25-50$, which have been employed
\citep{alb07a} for the modeling of the \object{Mrk 501} multi-wavelength
spectrum for the MAGIC 2005 observations. Note, however, that modeling the TeV $\gamma$-ray
emission from \object{PKS 2155-304}, in which also very short
timescale flares have been observed, seems to require Doppler
factors on the order of $\sim 100$ \citep{aha07}.

We can also infer a lower limit on the Doppler factor of the blob
by requiring the optical depth for $\gamma$-rays in the
synchrotron radiation of the blob to be below unity. For
another large flare observed from \object{Mrk 501},
\citet{bp99} estimated the optical depth for $\gamma$-rays
produced in the blob. Based on their Eq.~(6),
using the above mentioned parameters for the 2005 July 9 flare, we
estimate the Doppler factor of the blob to be $D_{\rm b}\ge 25$.
This rough estimate indicates that the blob has to move, at least
during a part of its path inside the jet, with Lorentz factors
above ten. Below we try to constrain the relation between the
Lorentz factors of the blob during its acceleration phase by using
the available information from observations in the context of the
SSC model considered.

In the discussion above it was assumed that the maximum energies
of the accelerated electrons are related to the magnetic field
strength at the blob location (see Eq.~\ref{eq3}),
because the synchrotron process constitutes the
dominant energy loss mechanism.
But they are also related
to the maximum energies of the $\gamma$-ray photons produced at
the respective specific location of the blob inside the jet (see
Eq.~\ref{eq5b}). In the considered model, the magnetic field in
the blob drops like $B\propto X^{-1} \propto R_{\rm b}^{-1}$, with the Doppler
factor of the blob $D_{\rm b}\propto R_{\rm b}$, and
$\gamma_{\rm e}\propto B^{-1/2}$. Therefore, the energies of the
observed $\gamma$-ray photons emitted at a specific location of
the blob should be related to the Doppler factor of the blob by
(see Eq.~\ref{eq5b})
\begin{eqnarray}
E_\gamma\propto D_{\rm b}^{3/2}.
\label{eq13}
\end{eqnarray}
\noindent The MAGIC collaboration reported the observation of a
time delay between $\gamma$-rays at different energies
$E_\gamma^{\rm min}$ and $E_\gamma^{\rm max}$. For the observed flare
of \object{Mrk 501} on 2005 July 9, these limiting values
differ by approximately one order of magnitude. Therefore, the
following relation should be fulfilled by the Doppler factors of
the blob at the two characteristic blob locations in the jet, which
correspond to the production of $\gamma$-rays with the energies
$E_\gamma^{\rm min}$ and $E_\gamma^{\rm max}$: $D_{\rm b}^{\rm
max}/D_{\rm b}^{\rm min}\approx \gamma_{\rm b}^{\rm
max}/\gamma_{\rm b}^{\rm min}\approx  10^{2/3}$.

As an example, based on the above constraints, we consider a blob
that is in its acceleration phase characterized by
Lorentz factors $\gamma_{\rm b}^{\rm min} = 10$ up to $\gamma_{\rm
b}^{\rm max} = 50$. For the reported time delay between the
flare at different energies ($\Delta\tau = 239 \,\mbox{s}$), we
can estimate (from Eq.~\ref{eq7}) the value of the parameter
$A\approx 1.7\times 10^{-4} \,\mbox{s}^{-1}$. The distance
traveled by the blob during the acceleration phase between the two
considered locations in the jet is $X_{\rm acc}\approx 7\times 10^{15}$ cm (cf.
Eq.~\ref{eq8}). Note that this distance scale is $\sim 20$ times
larger than the Schwarzschild radius of the $10^9 {\rm M}_\odot$
black hole \citep{fkt02,bhs03} expected in the center of
\object{Mrk 501}. Therefore, in our model, the acceleration region
of the blob containing relativistic electrons might well
be located relatively close to the central engine. At such
distances even the radiation field from the accretion disk around
the black hole might prevent the escape of $\gamma$-ray photons
\citep[see, e.g.,][]{bk95}. However, any conclusion on such
effects is limited by the unknown radiation fields provided by
the accretion disk and the broad line region in the case of BL~Lac
objects.

\section{Conclusions}
We have modified the classical homogeneous SSC model by allowing
the blob to accelerate during its propagation along the inner part
of the jet. The maximum energies of the electrons responsible for
the $\gamma$-ray production are determined by the synchrotron
process whose efficiency depends on the local strength of the
magnetic field in the blob. Therefore, the maximum energies of
$\gamma$-rays are also related to the acceleration phase of the
blob. In the resulting $\gamma$-ray flare, photons with lower
energies should be observed prior to those with higher energies.
Such a modified SSC model can explain the recently
observed arrival time difference of $\gamma$-rays belonging to
different energies in a flare of \object{Mrk 501}. In our model, for the first
time the MAGIC telescope was able to observe
the acceleration phase of a blob in the jet of \object{Mrk 501}.
Note that currently other models of the emission features at
VHE $\gamma$-rays \citep{alb07a,alb07b} also provide consistent
explanations.

Based on the measured time delay between the flares at different energies,
we estimate the distance
scale on which the acceleration of the blob occurred and predict
that the time delay should increase inversely proportional to the
energy of the $\gamma$-ray photons. For example, the delay between
the flare observed between $20 \,\mbox{GeV}$ and $200
\,\mbox{GeV}$ should be approximately an order of magnitude larger
than the observed delay between $200 \,\mbox{GeV}$ and $2
\,\mbox{TeV}$. This clear prediction can be tested by simultaneous
observations of strong flares at GeV and TeV $\gamma$-ray energies
using, e.g., the LAT instrument on board of GLAST (at GeV energies)
and Cherenkov telescopes (at TeV energies). For the specific case
of the observed \object{Mrk 501} flare, where a $\sim 240
\,\mbox{s}$ time delay has been reported between $\sim 190
\,\mbox{GeV}$ and $2.7 \,\mbox{TeV}$ (weighted average energies of
the reported energy ranges with a power-law slope of
$\alpha=-2.2$), we predict a corresponding $\sim 0.5-1 \,\mbox{h}$
delay between the peak position of this flare observed at $20
\,\mbox{GeV}$ and $200 \,\mbox{GeV}$. Observations of such delay
timescales are quantitatively consistent with the picture of an
accelerating blob within the jet.

\begin{acknowledgements}
This research is supported by the Polish MNiI grant 1P03D01028 and
by the Max Planck Society. R.M.W. acknowledges support by the DFG
cluster of excellence ``Origin and Structure of the Universe''.
\end{acknowledgements}

\bibliographystyle{aa}

%\listofobjects

\end{document}